\documentclass[aps,pre,preprint,nopreprintnumbers,amssymb,amsmath,showpacs,showkeys,fleqn,eqsecnum,superscriptaddress]{revtex4-1}

\usepackage[final]{graphicx}
\usepackage{amssymb}
\usepackage{amsmath}
\usepackage{dcolumn}

\begin{document}

\title{Fast Converging Path Integrals for Time-Dependent Potentials II: Generalization to Many-body Systems and Real-Time Formalism}
\author{Antun Bala\v{z}}\email[E-mail: ]{antun@ipb.ac.rs}
\affiliation{Scientific Computing Laboratory, Institute of Physics Belgrade, University of Belgrade, Pregrevica 118, 11080 Belgrade, Serbia}
\homepage[Home page: ]{http://www.scl.rs/}
\affiliation{Department of Physics, Faculty of Sciences, University of Novi Sad, Trg Dositeja Obradovi\' ca 4, 21000 Novi Sad, Serbia}
\author{Ivana Vidanovi\'c}
\author{Aleksandar Bogojevi\'c}
\author{Aleksandar Beli\'c}
\affiliation{Scientific Computing Laboratory, Institute of Physics Belgrade, University of Belgrade, Pregrevica 118, 11080 Belgrade, Serbia}
\homepage[Home page: ]{http://www.scl.rs/}
\author{Axel Pelster}
\affiliation{Fachbereich Physik, Universit\" at Duisburg-Essen, Lotharstra\ss e 1, 47048 Duisburg, Germany}

\begin{abstract}
Based on a previously developed recursive approach for calculating the short-time expansion of the propagator for systems with time-independent potentials and its time-dependent generalization for simple single-particle systems, in this paper we present a full extension of this formalism to a general quantum system with many degrees of freedom in a time-dependent potential. Furthermore, we also present a recursive approach for the velocity-independent part of the effective potential, which is necessary for calculating diagonal amplitudes and partition functions, as well as an extension from the imaginary-time formalism to the real-time one, which enables to study the dynamical properties of quantum systems. The recursive approach developed here allows an analytic derivation of the short-time expansion to orders that have not been accessible before, using the implemented SPEEDUP symbolic calculation code. The analytically derived results are extensively numerically verified by treating several models in both imaginary and real time.
\end{abstract}

\maketitle

\section{Introduction}
\label{sec:intro}
Although a large number of physical systems admit studies of their basic properties using different types of time-independent formalisms, in many important cases one has to explicitly take into account the time dependence and to apply the appropriate approach, i.e.~one of available time-dependent analytical or numerical methods. This has to be done even for systems with time-independent potentials if their dynamical properties and time evolution is studied. However, for describing systems in genuinely time-dependent external potentials or for rotating systems, using of such approaches is always necessary. This applies equally to classical and quantum systems, and various methods have been developed to address relevant physical problems. For quantum systems a number of general methods is available, ranging from time-dependent perturbation theory and variational perturbation theory, to specialized approaches such as the Density Matrix Renormalization Group (DMRG) \cite{schollwoeck, hallberg}, 
the Density Functional Theory (DFT) \cite{runge, onida}, and the Density Matrix Functional Theory (DMFT) \cite{pernal}.

In addition to these generic schemes, several specific numerical methods have been developed for enabling a quantitative description of quantum systems that have attracted a significant research interest. This includes studies of ultra-cold quantum gases \cite{dalibard2004, fetter2005, abdullaevpla, pelster1,theodorakis, alon, pelster2, mason2009, aftalion1,aftalion2} and optical lattices \cite{morsch, inguscio, ketterle1, ketterle2}, whose comprehensive and highly tunable features make them an important example of Feynman's quantum simulator \cite{qs}. In such numerical approaches \cite{chinkrotscheck, hernandez, ciftja, sakkos} usually a second-order algorithm in the propagation time is used, which is basically the same as in the time-independent case. 
However, also higher order schemes have been derived, including fourth \cite{bandrauk, blanes, chinchen, bayepla} and higher-order expansions \cite{omelyanpre, omelyan, bayepre, zillich}.

The main goal of this paper is to develop a formalism which enables a systematic improvement in the convergence order of numerical algorithms for general time-dependent quantum systems.
In our previous paper \cite{fcpitdp1}, we have extended the earlier established approach \cite{prl-speedup, prb-speedup, pla-manybody, balazpre} for obtaining a high-order short-time expansion of transition amplitudes for time-independent potentials to the important time-dependent case. In that paper, we have first calculated a short-time expansion of a generic transition amplitude using the path integral formalism to fourth order in the propagation time, which has then served the important purpose: to introduce the proper functional form of the ansatz for the ideal discretized effective potential, as well as its short-time double series expansion in both the propagation time and the discretized velocity. Using both the forward and backward Schr\" odinger equation for transition amplitudes, we have then derived the appropriate equation for the ideal effective potential, which represents an important generalization of the equation derived in Ref.~\cite{balazpre} for the time-independent case. Using the double series expansion ansatz for the effective potential, we have been able to set up and solve recursive relations for the effective potential for one-dimensional systems, and to numerically verify that higher-order analytic approximative results for transition amplitudes obtained in this way, indeed, give the correct convergence order for a number of models.

In this paper we further develop and generalize the time-dependent approach introduced in Ref.~\cite{fcpitdp1}. First, in Sec.~\ref{sec:eff} we briefly review the time-dependent effective action approach in the path-integral formalism of quantum mechanics \cite{feynman,feynmanhibbs,feynmanstat,kleinertbook}, as well as the main results of our previous paper \cite{fcpitdp1}. Then we extend the recursive approach for calculating the short-time expansion of the effective potential to quantum systems with many degrees of freedom in Sec.~\ref{sec:mb}. In this section, the generalized approach is also numerically verified for the case of a simple time-dependent multi-component harmonic oscillator system, and possible relevant physical applications in the realm of ultracold quantum gases are briefly indicated. In Sec.~\ref{sec:1dZ} we present another important extension of the time-dependent formalism, and set up a specific recursive relation for calculating diagonal transition amplitudes, which are necessary for a numeric high-precision calculation of partition functions. In this case, a simplified set of recursive relations is obtained, which is numerically verified for several model potentials. Finally, Sec.~\ref{sec:rt} illustrates how the developed imaginary-time formalism can be transformed into a real-time one, and its applicability is numerically demonstrated by treating several models. We also show how the real-time evolution can be described using the time-dependent effective action approach, and how the associated numerical errors can be assessed and controlled in typical applications.

\section{Effective Action Approach for Systems with Time-Depen\-dent Potentials}
\label{sec:eff}
In this section we give a brief overview of the time-dependent effective action approach established in Ref.~\cite{fcpitdp1}. This formalism considers a non-relativistic quantum multi-component system in $d$ spatial dimensions with a Hamiltonian of the form
\begin{equation}
\label{eq:hamiltonian}
\hat H(\hat{\mathbf p}, \hat{\mathbf q}, t)=\sum_{i=1}^P\frac{\hat{\mathbf p}_{(i)}^2}{2M_{(i)}}+V(\hat{\mathbf q}, t)\, ,
\end{equation}
where $P$ denotes the number of particles in the system and the $P\times d$ dimensional vectors $\mathbf q$ and $\mathbf p$ contain positions and momenta of all particles, while the parenthetic subscript 
$(i)$ denotes the particle number. For such a system we consider the calculation of the transition amplitudes
\begin{equation}
\label{eq:AU}
A(\mathbf a,t_a; \mathbf b, t_b)=\langle \mathbf b, t_b|\hat U(t_a\to t_b)|\mathbf a, t_a\rangle\, ,
\end{equation}
where the vectors $\mathbf a$ and $\mathbf b$ describe the positions of all particles at the initial and final time $t_a$ and $t_b$, respectively. The above transition amplitude is a coordinate-space matrix element of the evolution operator, which describes the propagation of the system (\ref{eq:hamiltonian}) from $t_a$ to $t_b$, and is defined by the time-ordered exponential
\begin{equation}
\label{eq:UT}
\hat U(t_a\to t_b)=\hat{T}\exp\left\{-\frac{i}{\hbar}\int_{t_a}^{t_b}dt\, \hat H(\hat{\mathbf p}, \hat{\mathbf q}, t) \right\}\, .
\end{equation}
In the path-integral formalism, the transition amplitude can be expressed by a coordinate-space path integral
\begin{equation}
\label{eq:pi}
A(\mathbf a,t_a; \mathbf b, t_b)=\int_{\mathbf q(t_a)=\mathbf a}^{\mathbf q(t_b)=\mathbf b}  {\cal D} \mathbf q(t)\, \exp \left\{\frac{i}{\hbar} S[\mathbf q]\right\}\, ,
\end{equation}
where the integration is defined over all possible trajectories $\mathbf q(t)$. This usually involves the discretization of the trajectories, which is usually performed by dividing the time evolution from $t_a$ to $t_b$ into $N$ equal time steps. In the above equation, $S$ denotes the action for a given trajectory $\mathbf q(t)$:
\begin{equation}
\label{eq:Sdef}
S[\mathbf q]=\int_{t_a}^{t_b} dt\, \left\{ \frac{1}{2}\dot{\mathbf q}^2(t)-V(\mathbf q(t), t)\right\}\, .
\end{equation}
The common step at this point is to switch to the imaginary-time formalism, which is usually applied in numerical simulations \cite{ceperley}, due to problems which may arise from the oscillatory nature of the integrand in the real-time approach. We will do so as well in the next two sections, but in Sec.~\ref{sec:rt} we will switch the developed formalism back to the real time and demonstrate how it can be used for studying the dynamics and real-time evolution of quantum systems. After Wick rotation to the imaginary time, the transition amplitude in the path-integral formalism is expressed as
\begin{equation}
\label{eq:piit}
A(\mathbf a,t_a; \mathbf b, t_b)=\int_{\mathbf q(t_a)=\mathbf a}^{\mathbf q(t_b)=\mathbf b}  {\cal D} \mathbf q(t) \, e^{-\frac{1}{\hbar} S_E[\mathbf q]}\, ,
\end{equation}
where the Minkowski action is now replaced by its imaginary-time counterpart, the Euclidean action
\begin{equation}
\label{eq:SEdef}
S_{\rm E}[\mathbf q]=\int_{t_a}^{t_b} dt\, \left\{ \frac{1}{2}\dot{\mathbf q}^2(t)+V(\mathbf q(t), t)\right\}\, ,
\end{equation}
which is actually the energy functional for the system.

As was shown previously for time-independent potentials \cite{balazpre}, as well as for the time-dependent ones in Ref.~\cite{fcpitdp1}, the exact imaginary-time transition amplitudes can be expressed in the form
\begin{equation}
\label{eq:Aexacta}
A(\mathbf a, t_a; \mathbf b,t_b)=\frac{1}{(2\pi\varepsilon)^{Pd/2}}\ e^{-  S^*(\mathbf x, \bar{\mathbf x}; \varepsilon, \tau)}\, ,
\end{equation}
where $S^*$ stands for the ideal discretized action and depends on the coordinate mid-point $\mathbf x=(\mathbf a+\mathbf b)/2$, the discretized velocity $\bar{\mathbf x}=(\mathbf b-\mathbf a)/2$, the time interval $\varepsilon=t_b-t_a$, and the time mid-point $\tau=(t_a+t_b)/2$. Note that we have used the convention $\hbar=1$, and we have restricted ourselves to particles with unitary masses $m_{(i)}=1$. The ideal discretized action further reads \cite{fcpitdp1, pla-euler, pre-ideal}
\begin{equation}
\label{eq:Aexactb}
S^*(\mathbf x, \bar{\mathbf x}; \varepsilon, \tau)
= \frac{2}{\varepsilon}\bar{\mathbf x}^2+\varepsilon W(\mathbf x, \bar{\mathbf x}; \varepsilon, \tau)\, ,
\end{equation}
where $W$ is the ideal discretized effective potential, which also depends on the time mid-point $\tau$ due to the explicit time dependence of the potential $V$. This represents the major difference in the formalism compared to the previously developed one in Ref.~\cite{balazpre} for the time-independent case.

In order to determine a partial differential equation for the ideal effective potential $W$, we have derived in Sec.~4 of our previous paper \cite{fcpitdp1} the forward and the backward Schr\"odinger equation for time derivatives of the transition amplitude with respect to the initial and final time, and have found that they obey
\begin{eqnarray}
\label{eq:dtbA}
&&\partial_{t_b}\, A(\mathbf a,t_a; \mathbf b, t_b) =- \hat H_b\, A(\mathbf a,t_a; \mathbf b, t_b)\, ,\\
\label{eq:dtaA}
&&\partial_{t_a}\, A(\mathbf a,t_a; \mathbf b, t_b) = \hat H_a\, A(\mathbf a,t_a; \mathbf b, t_b)\, ,
\end{eqnarray}
where $\hat H_b$ stands for the coordinate-space Hamilton operator $\hat H_b=H(-i \boldsymbol{\partial}_{\mathbf b}, \mathbf b, t_b)$, in which momentum and position operators 
are replaced by their coordinate-space representations at $\mathbf b$ (and similarly for $\hat H_a$). When we insert the ansatz (\ref{eq:Aexacta}) together with (\ref{eq:Aexactb}) into the equations (\ref{eq:dtbA}) and (\ref{eq:dtaA}) , we obtain the corresponding partial differential equation for the effective potential in the form
\begin{equation}
W+\bar{\mathbf x}\cdot\bar{\boldsymbol{\partial}}\,W+\varepsilon\partial_\varepsilon W
-\frac{1}{8}\,\varepsilon\partial^2 W-\frac{1}{8}\,\varepsilon\bar\partial^2 W
+\frac{1}{8}\,\varepsilon^2(\boldsymbol{\partial} W)^2
+\frac{1}{8}\,\varepsilon^2(\bar{\boldsymbol{\partial}} W)^2=\frac{1}{2}\, (V_++V_-)\, ,
\label{eq:Weq}
\end{equation}
where $V_\pm=V(\mathbf x \pm \bar{\mathbf x}, \tau\pm\frac{\varepsilon}{2})$, i.e. $V_-=V(\mathbf a, t_a)$ and $V_+=V(\mathbf b, t_b)$. As expected, the form of this equation is the same as Eq.~(29) from Ref.~\cite{balazpre}, and if the potential $V$ does not depend on time, we immediately obtain the previously derived result.

The above equation is the starting point for calculating the effective potential. Naturally, it can be solved analytically only for exactly solvable models. However, if the propagation time is short, we can perform a short-time expansion of the effective potential and set up appropriate equations for the coefficients in this expansion. In our previous paper \cite{fcpitdp1} this was done for one-dimensional systems, and it was shown that we have to use a double expansion of the effective potential in both $\varepsilon$ and $\bar{\mathbf x}$. The reason for this is the fact that the propagation in the imaginary time is equivalent to diffusion, and therefore, on the average, we expect the diffusion relation $\bar{\mathbf x}^2\sim \varepsilon$ to hold. This allows us to effectively couple the two expansion parameters and to establish a unique counting of powers in $\varepsilon$ for all terms in the expansion for $W$. However, if the diffusion relation is not applicable, the two expansions can be considered as independent, and the whole approach can still be applied, with an independent counting of powers in $\varepsilon$ and in $\bar{\mathbf x}$.

In the next section we will analytically derive a systematic short-time expansion of the effective potential $W$ for quantum many-body systems, which yields a significant improvement in the convergence of numerically calculated transition amplitudes and partition functions for systems in time-dependent potentials. As we see from Eq.~(\ref{eq:Aexactb}), if the effective potential $W$ is calculated to order $\varepsilon^{p-1}$, we get the effective action correct to order $\varepsilon^p$. If we take into account the normalization factor in the expression (\ref{eq:Aexacta}), the corresponding error in the calculation of the short-time amplitude is given by
\begin{equation}
\label{eq:Apscaling}
A_p(\mathbf a, t_a; \mathbf b,t_b)=A(\mathbf a, t_a; \mathbf b,t_b)+O(\varepsilon^{p+1-Pd/2})\, ,
\end{equation}
where subscript $p$ denotes that we use the effective action of that order in $\varepsilon$. Therefore, we see that the analytical calculation of higher-order effective actions is beneficial, since it provides an analytic approximation for transition amplitudes which yield high-precision results of the desired order in numerical calculations.

\section{Multi-component systems}
\label{sec:mb}

Now we focus on the development of the recursive formalism for calculating the effective potential for the case of a general multi-component non-relativistic quantum system of $P$ particles in $d$
dimensions by extending the one-dimensional calculations of Sec.~5 in our previous paper \cite{fcpitdp1}. Note that such a formalism is needed even for studies of single-particle systems in two or three spatial dimensions, which have more than one degree of freedom. For example, such time-independent many-body effective actions of level $p=21$ have already been used for a numerical study of fast-rotating Bose-Einstein condensates \cite{becpla}, as well as a high-precision calculation of the energy spectra and eigenfunctions of several two-dimensional models \cite{diag1, diag2}. The presented extension of the many-body formalism will allow studies of such systems in external time-dependent potentials, as well as the investigation of the formation and evolution of vortices \cite{aftalion1, aftalion2} and other dynamical phenomena. We also plan to study collective oscillation modes of Bose-Einstein condensates with a parametrically 
modulated interaction \cite{bagnato, abdullaev}.

To develop the time-dependent many-body formalism, we solve the partial differential equation (\ref{eq:Weq}) for the effective potential $W$ by using a multi-dimensional many-particle generalization of the double power expansion used in Eq.~(33) of our previous paper \cite{fcpitdp1} for one-component systems, which has the form
\begin{equation}
\label{eq:MBWdouble}
W(\mathbf x,\bar{\mathbf x};\varepsilon, \tau)=\sum_{m=0}^{\infty}\sum_{k=0}^{m}\Big\{W_{m,k}(\mathbf x, \bar{\mathbf x}; \tau)\,\varepsilon^{m-k}
+ W_{m+1/2,k}(\mathbf x, \bar{\mathbf x}; \tau)\,\varepsilon^{m-k}\Big\} \, .
\end{equation}
Here we have introduced the contractions
\begin{equation}
\label{eq:MBWevencontr}
W_{m, k}(\mathbf x,\bar{\mathbf x}; \tau)=\bar x_{i_1}\cdots \bar x_{i_{2k}}\,  c_{m, k}^{i_1,\ldots i_{2k}}(\mathbf x; \tau)\, ,
\end{equation}
\begin{equation}
\label{eq:MBWoddcontr}
W_{m+1/2, k}(\mathbf x,\bar{\mathbf x}; \tau)=\bar x_{i_1}\cdots \bar x_{i_{2k+1}}\, c_{m+1/2, k}^{i_1,\ldots i_{2k+1}}(\mathbf x; \tau)\, ,
\end{equation}
in such a way that they correspond to
the case of the time-independent potential \cite{balazpre}. In the above relations we assume the Einstein convention that summation 
over repeated indices is performed. Introducing such contractions of tensorial 
coefficients $c$ significantly simplifies in this case
the analytic derivation and also provides a key ingredient for implementing the many-body recursion relations in symbolic calculations using e.g.
Mathematica software package \cite{mathematica}. Otherwise, the task to explicitly symmetrize the coefficients would amount to a complexity of the algorithm which scales 
with the number of possible permutations $(P\times d)!$ and which would not be feasible even for a very moderate number of particles. Using scalar quantities, which are
obtained by contracting the coefficients with the 
discretized velocity $\bar{\mathbf x}$, efficiently solves this problem.

The multiple-component version of the expression from the right-hand side of Eq.~(\ref{eq:Weq}) has the form
\begin{eqnarray}
&&\hspace*{-13mm}\frac{1}{2}\, (V_++V_-)=\nonumber\\
&&\hspace*{-8mm}\sum_{m=0}^\infty\sum_{k=0}^m\left\{ \frac{\Pi(m, k)\, \varepsilon^{m-k}}{(2k)!\, (m-k)!\, 2^{m-k}}(\bar{\mathbf x}\cdot\boldsymbol{\partial})^{2k}
+\frac{(1-\Pi(m, k))\,  \varepsilon^{m-k}}{(2k+1)!\, (m-k)!\, 2^{m-k}} (\bar{\mathbf x}\cdot\boldsymbol{\partial})^{2k+1} \right\} \stackrel{(m-k)}{V} ,
\label{eq:MBVpmexp}
\end{eqnarray}
where $\stackrel{(m-k)}{V}$ represents $(m-k)$-th partial derivative of the potential with respect to the time $\tau$. After inserting the above expression into the partial differential equation (\ref{eq:Weq}) for the effective potential we straightforwardly obtain recursive relations for even- and odd-power contractions $W_{m, k}$ and $W_{m+1/2, k}$:
\begin{eqnarray}
&&\hspace*{-10mm}
8(m+k+1)\, W_{m,k}=8 \frac{\Pi(m, k)\, (\bar{\mathbf x}\cdot\boldsymbol{\partial})^{2k} \stackrel{(m-k)}{V}}{(2k)!\, (m-k)!\, 2^{m-k}} + \bar{\partial}^2\, W_{m,k+1}
+\partial^2\, W_{m-1,k}\nonumber\\
&&\hspace*{2mm}-\sum_{l, r}\Big\{\boldsymbol{\partial} W_{l,r}\cdot \boldsymbol{\partial} W_{m-l-2,k-r}
+\boldsymbol{\partial} W_{l+1/2,r} \cdot \boldsymbol{\partial} W_{m-l-5/2,k-r-1}\nonumber\\
\label{eq:receven}
&&\hspace*{2mm}
+\bar{\boldsymbol{\partial}} W_{l,r} \cdot \bar{\boldsymbol{\partial}} W_{m-l-1,k-r+1} 
+ \bar{\boldsymbol{\partial}} W_{l+1/2,r} \cdot \bar{\boldsymbol{\partial}} W_{m-l-3/2,k-r}\Big\}\, ,\\
&&\hspace*{-10mm}
8(m+k+2)\, W_{m+1/2,k}=8 \frac{(1-\Pi(m, k))\, (\bar{\mathbf x}\cdot\boldsymbol{\partial})^{2k+1} \stackrel{(m-k)}{V}}{(2k+1)!\, (m-k)!\, 2^{m-k}}
+ \bar{\partial}^2\, W_{m+1/2,k+1}+ \partial^2\, W_{m-1/2,k}\nonumber\\
&&\hspace*{2mm}
-\sum_{l, r}\Big\{  \boldsymbol{\partial} W_{l,r}\cdot \boldsymbol{\partial} W_{m-l-3/2,k-r}
+ \boldsymbol{\partial} W_{l+1/2,r} \cdot \boldsymbol{\partial} W_{m-l-2,k-r}\nonumber\\
&&\hspace*{2mm}
+ \bar{\boldsymbol{\partial}} W_{l+1/2,r} \cdot \bar{\boldsymbol{\partial}} W_{m-l-1,k-r+1}
+ \bar{\boldsymbol{\partial}} W_{l,r} \cdot \bar{\boldsymbol{\partial}} W_{m-l-1/2,k-r+1}\Big\}\, .
\label{eq:recodd}
\end{eqnarray}
The diagonal contractions can easily be calculated in a closed form as in the single-particle one-dimensional case, yielding
\begin{eqnarray}
\label{eq:Wmm}
W_{m,m}&=&\frac{1}{(2m+1)!}(\bar{\mathbf x}\cdot \boldsymbol{\partial})^{2m}\,V\, ,\\
\label{eq:Wm12m}
W_{m+1/2, m}&=&0\, .
\end{eqnarray}
Thus, the recursion relations (\ref{eq:receven}) and (\ref{eq:recodd}) can be solved up to a given order $p$ together with  (\ref{eq:Wmm}) and (\ref{eq:Wm12m})
by using a similar procedure as before.  Here we give the solution up to order $p=4$, which generalizes the 
previously given solution for the simple  case $P=d=1$, obtained in Ref.~\cite{fcpitdp1}. For $m=0$ we only have the naive $p=1$ term, i.e.
\begin{equation}
\label{nai}
W_{0,0}=V\, ,
\end{equation}
while $m=1$ yields the first non-trivial even-power terms
\begin{eqnarray}
W_{1,1}&=&\frac{1}{6}\,(\bar{\mathbf x}\cdot\boldsymbol{\partial})^2 V\, ,\\
W_{1,0}&=&\frac{1}{12}\partial^2\, V\, ,
\end{eqnarray}
which are
sufficient to construct $p=2$ effective action. The next term we calculate is the odd-power contraction for $m=1$, i.e.
\begin{equation}
W_{3/2, 0}= \frac{1}{6}\,(\bar{\mathbf x}\cdot\boldsymbol{\partial}) \dot{V}\, ,
\end{equation}
which contains the explicit time derivative of $V$. 
For $m=2$ we obtain the next order of even-power terms:
\begin{eqnarray}
W_{2,2}&=&\frac{1}{120}\,(\bar{\mathbf x}\cdot\boldsymbol{\partial})^4 V\, ,\\
W_{2,1}&=&\frac{1}{120}\,(\bar{\mathbf x}\cdot\boldsymbol{\partial})^2 \partial^2\, V\, ,\\
W_{2, 0}&=&\frac{1}{24}\ddot{V}+\frac{1}{240}\,\partial^4\, V -\frac{1}{24}\, (\boldsymbol{\partial} V)^2\, .
\end{eqnarray}
These terms, together with the previously calculated ones, are sufficient to construct level $p=3$ effective action. In order to be able to complete $p=4$ effective action derivation, we still need to 
calculate odd-power contractions corresponding to $m=2$
\begin{eqnarray}
W_{5/2,1}&=&\frac{1}{60}\,(\bar{\mathbf x}\cdot\boldsymbol{\partial})^3\, \dot{V}\, ,\\
W_{5/2,0}&=&\frac{1}{120}\,(\bar{\mathbf x}\cdot\boldsymbol{\partial})\, \partial^2 \dot{V}\, ,\\
\end{eqnarray}
as well as even-power contractions for $m=3$,
\begin{eqnarray}
&&\hspace*{-2mm}W_{3,3}=\frac{1}{5040}\,(\bar{\mathbf x}\cdot\boldsymbol{\partial})^6 V\, ,\\
&&\hspace*{-2mm}W_{3,2}=\frac{1}{3360}\,(\bar{\mathbf x}\cdot\boldsymbol{\partial})^4 \partial^2\, V\, ,\\
&&\hspace*{-2mm}W_{3, 1}=\frac{1}{3360}\,(\bar{\mathbf x}\cdot\boldsymbol{\partial})^2 \partial^4\, V+ \frac{1}{80}\,(\bar{\mathbf x}\cdot\boldsymbol{\partial})^2\, \ddot{V}
-\frac{1}{360}\, \Big( (\bar{\mathbf x}\cdot\boldsymbol{\partial}) \boldsymbol{\partial} V\Big)^2\nonumber\\
&&\hspace*{12mm} -\frac{1}{120}\, (\boldsymbol{\partial} V)\cdot (\bar{\mathbf x}\cdot
\boldsymbol{\partial})^2 \boldsymbol{\partial} V\, ,\\
&&\hspace*{-2mm}W_{3, 0}=\frac{1}{6720}\,\partial^6\, V+  \frac{1}{480}\,\partial^2\, \ddot{V}-\frac{1}{360}\, (\partial_i\boldsymbol{\partial} V)\cdot (\partial_i\boldsymbol{\partial} V)
-\frac{1}{120}\, (\boldsymbol{\partial} V)\cdot \partial^2 \boldsymbol{\partial} V\, . \label{mbb}
\end{eqnarray}
This concludes the calculation of level $p=4$ effective action for a general many-body quantum system. As before, the obtained results automatically reduce to the already known effective actions for 
the time-independent potentials \cite{balazpre} if we set all time-derivatives of the potential to zero. Furthermore, the many-body results (\ref{nai})--(\ref{mbb}) reduce for the special
case $P=d=1$ to the previous time-dependent results \cite{fcpitdp1}. The outlined procedure continues in the same way for higher levels $p$. We have automatized this procedure and implemented it in our SPEEDUP code \cite{speedup} using the Mathematica software package \cite{mathematica} for symbolic calculus.

\begin{figure}[!t]
\centering
\includegraphics[width=8cm]{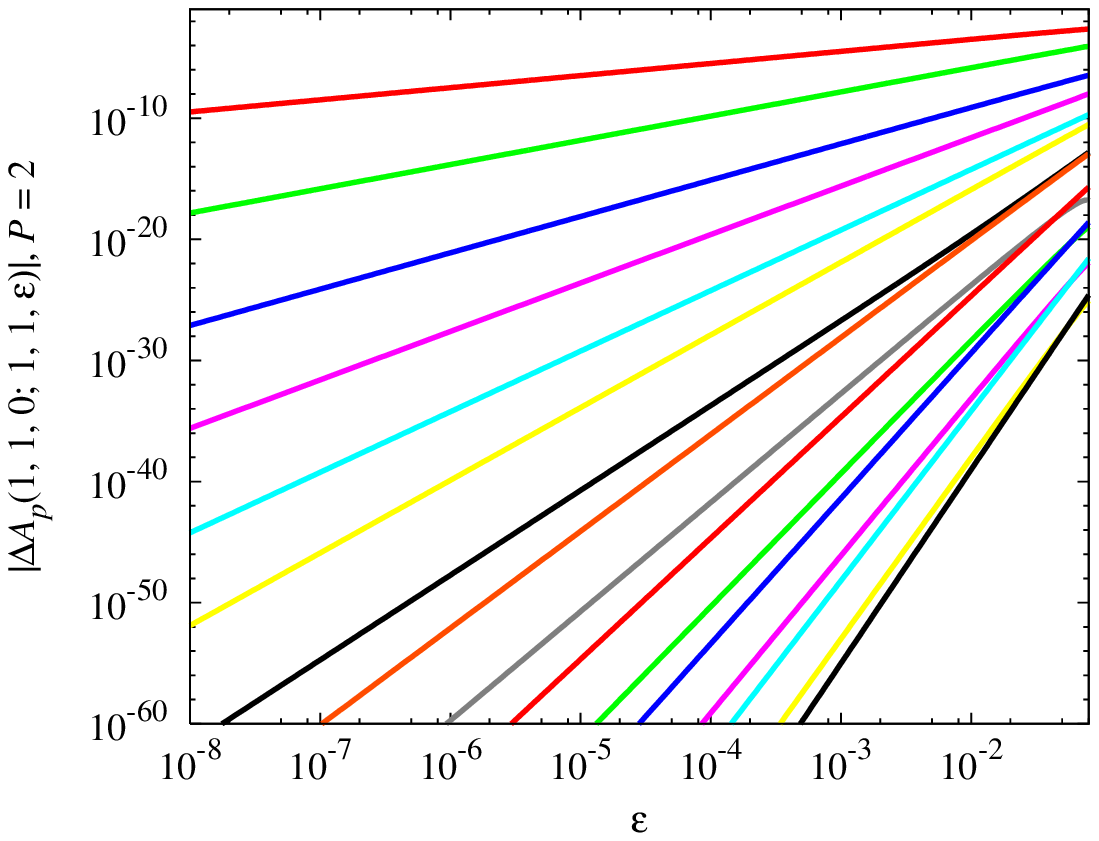}
\includegraphics[width=8cm]{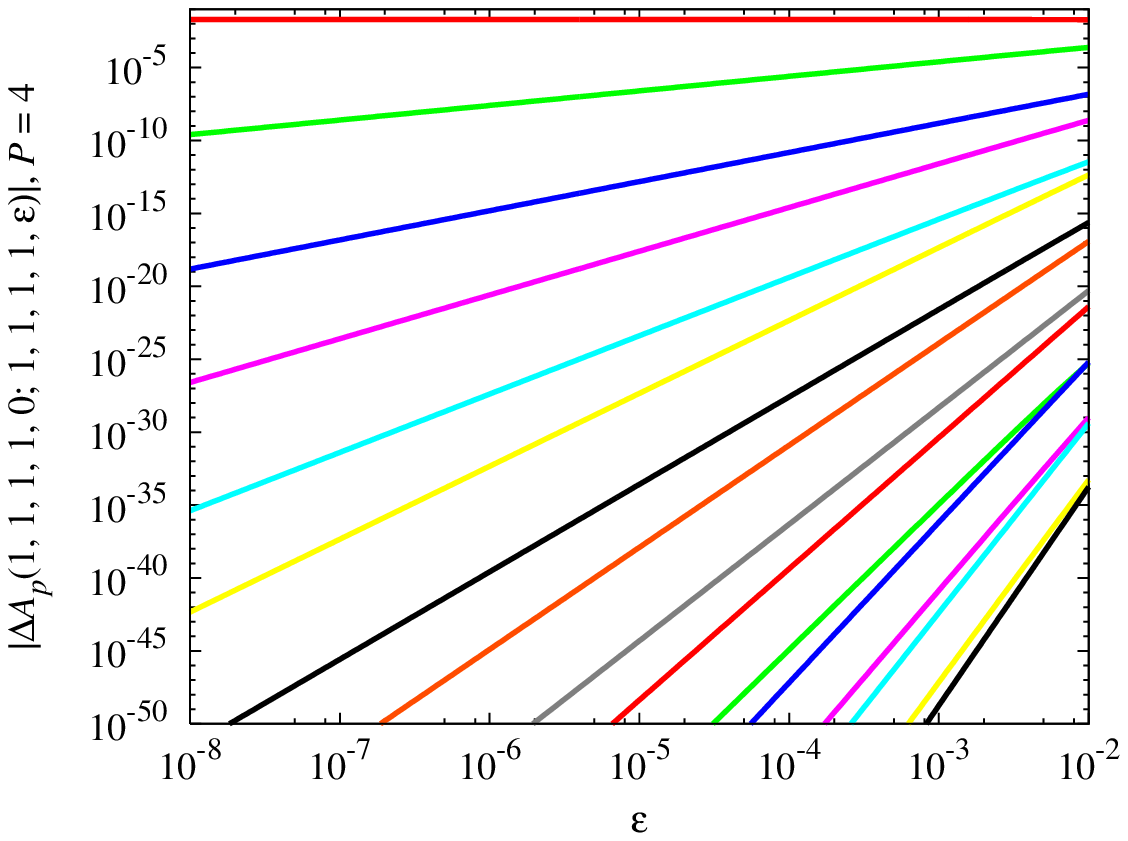}
\includegraphics[width=8cm]{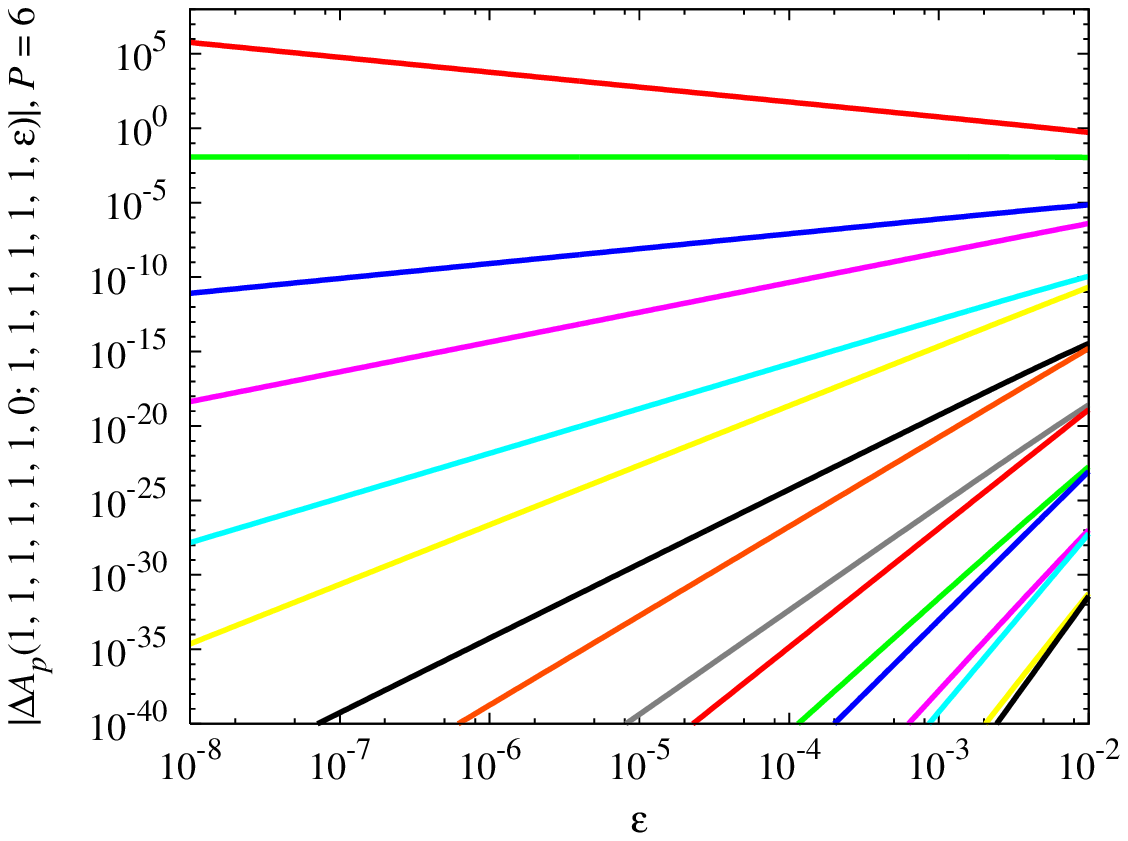}
\caption{Deviations of diagonal transition amplitudes $|\Delta A_p(\mathbf{a}=\mathbf{1}, t_a=0; \mathbf{b}=\mathbf{1}, t_b=\varepsilon)|$ from the exact 
values as a function of propagation time $\varepsilon$ for a multi-component system  (\ref{eq:mctdho}) of time-dependent harmonic oscillators: (top left) $P=2$ oscillators, 
with $\omega_1^2(t)=1+\frac{1}{2}\, \sin^2 2t$, $\omega_2^2(t)=1+\frac{1}{2}\, \cos 2t$; (top right) $P=4$ oscillators, with $\omega_1^2(t)$, $\omega_2^2(t)$, $\omega_3^2(t)=2
+\cos 5t$, $\omega_4^2(t)=4+\sin^2 4t$; (bottom)  $P=6$ oscillators, with $\omega_1^2(t)$, $\omega_2^2(t)$, $\omega_3^2(t)$, $\omega_4^2(t)$, $\omega_5^2(t)=2+\sin^2 t$, 
$\omega_6^2(t)=4+2\cos 3t$. Each plot gives results for transition amplitudes calculated using the effective action levels $p=1, 2, \ldots, 16$, from top to bottom.}
\label{fig:mc}
\end{figure}

In order to numerically verify the developed expressions for the case of multi-component quantum systems, we will calculate transition amplitudes of a set of time-dependent harmonic oscillators,
\begin{equation}
\label{eq:mctdho}
V(\mathbf{q}, t)=\sum_{i=1}^P\frac{1}{2}\, \omega_i^2(t)\, q_i^2\, ,
\end{equation}
which represents the archetypical model for many physical phenomena. This exactly solvable model allows us to compare analytical expressions obtained from recursive relations 
and to verify that using level $p$ effective action leads to values of transition
amplitudes which are correct up to order $\varepsilon^{p+1-Pd/2}$. As we can see in Fig.~\ref{fig:mc} for a system 
of $P=2, 4, 6$ time-dependent oscillators, the respective 
scaling is perfect. The middle and bottom plots illustrate another important feature of the short-time expansion for 
multi-component systems: as the number of components of the system $Pd$ increases, the exponent $p+1-Pd/2$ may become zero or negative for a given effective action level $p$. 
This leads to the peculiar behavior observed in the middle and bottom plots for small values of $p$, with the deviation of the amplitude being constant ($P=4$, $p=1$ in the 
middle plot, $P=6$, $p=2$ in the bottom plot) or even increasing ($P=6$, $p=1$ in the bottom plot) when $\varepsilon$ is decreased. Thus, this prevents the calculation of the 
transition amplitude 
with high accuracy, which is, in principle, expected to be possible by decreasing $\varepsilon$. As we see, to enable such high-accuracy calculations, one has to use an effective 
action with sufficiently high level $p$. The important contribution of the presented approach lies in the fact that it offers a systematic formalism for deriving such higher 
order expressions for a general quantum multi-component system.

As in the case of single-particle one-dimensional systems, computational complexity of higher-order effective actions might significantly increase for higher values of $p$ (typically exponentially, see e.g. Fig.~5 in Ref.~\cite{fcpitdp1}). This increase strongly depends on the type of potential, but we expect to obtain significant computational benefits at least for moderate values of $p$. The optimal value of the level $p$ to be used in practical applications can always be found through a small-scale numerical complexity study, by measuring relative increase in CPU times as a function of $p$. Using this data, as well as the known scaling of Monte Carlo errors and CPU times on the time step and number of samples, optimal value of $p$ is obtained by minimizing the CPU time for a required precision of transition amplitudes.

At the end of this section, we emphasize that the obtained discretized effective actions can be used for solving a plethora of non-equilibrium many-body quantum problems within the exact diagonalization \cite{diag1, diag2} or Path Integral Monte Carlo approach, including the continuous-space worm algorithm \cite{boninsegni}. For instance, in typical experimental setups with ultracold quantum gases harmonic
or anharmonic confining potentials are generically switched on and off, thus generating natural non-equilibrium situations. As so far mainly quenched potentials have been considered,
it would certainly be rewarding to study in a systematic way how the time scale, upon which a potential is switched off, influences the observed time-of-flight absorption pictures.
Another upcoming research field is the investigation of the phenomenon of parametric resonance in Bose-Einstein condensates. A first experiment, where the s-wave scattering length of
$^7$Li atoms has been modulated periodically with the help of a Feshbach resonance, has recently been performed \cite{bagnato}. In order to understand the observed resonance spectrum both
analytical methods from nonlinear dynamics \cite{abdullaev} and numerical methods as the presented fast converging path-integral approach have to be combined.

\section{Velocity-independent Part of the Effective Potential}
\label{sec:1dZ}

Now we turn our attention to the special case of the velocity independent part of the effective potential. It determines the diagonal amplitudes, for which the discretized velocity $\bar{\mathbf x}$ is equal to zero. The efficient and precise calculation of diagonal transition amplitudes is essential for many quantum statistical problems, since it provides a direct method to obtain partition functions, and can be used to calculate energy spectra, density profiles, and other relevant physical quantities. Therefore, we will derive a new set of recursive relations for the coefficients which determine the velocity-independent part of the effective potential. For simplicity, we will present the derivation for one-dimensional systems, where the effective potential for $\bar x=0$ can be written as
\begin{equation}
\label{eq:W0double}
W_0(x;\varepsilon, \tau) \equiv W (x,0;\varepsilon, \tau)=\sum_{m=0}^{\infty}c_{m,0}(x, \tau)\,\varepsilon^{m}\, .
\end{equation}
Such recursive relations for coefficients $c_m\equiv c_{m,0}$ will turn out to be much simpler than the full set of recursions obtained in the previous paper \cite{fcpitdp1}.

In order to derive equations determining the coefficients $c_m$, we have to perform the limit $\bar x\to 0$ in the partial differential equation (\ref{eq:Weq}) 
for the effective potential $W$. This is nontrivial, since the equation 
contains derivatives with respect to $\bar x$. Therefore, we have to re-examine both Schr\"odinger equations (\ref{eq:dtbA}) and (\ref{eq:dtaA}), and express them using the variables $x$, $\bar x$, $\varepsilon$, and $\tau$. After a change of variables, we get the following system of equations for the transition amplitude:
\begin{eqnarray}
\label{eq:depsxbarxA}
\left[\partial_\varepsilon-\frac{1}{8}\,\partial^2-\frac{1}{8}\,\bar\partial^2
+\frac{1}{2}\, (V_++V_-)\right]A(x, \bar{ x}; \varepsilon, \tau)&=&0\, ,\\
\label{eq:dt}
\left[\partial_\tau -\frac{1}{2}\partial\bar\partial+V_+ -V_- \right]A( x, \bar{ x}; \varepsilon, \tau) &=& 0\, .
\end{eqnarray}
If we take the derivative with respect to $\bar x$ of the second equation, use it to express the term $\partial\bar\partial^2 A$, and insert it into the derivative of the first equation with 
respect to $x$, we obtain the partial differential equation
\begin{equation}
\label{eq:A0comb}
\hspace*{-10mm}
\partial_\varepsilon\partial A-\frac{1}{8}\partial^3 A-\frac{1}{4}\partial_\tau\bar\partial A-\frac{1}{4}A\, \bar\partial(V_+-V_-)
-\frac{1}{4}(V_+-V_-)\, \bar\partial A+ \frac{1}{2}A\, \partial(V_++V_-)
+\frac{1}{2}(V_++V_-)\, \partial A=0\, ,
\end{equation}
in which it is easier to perform the required $\bar x\to 0$ limit. In the terms that do not contain derivatives with respect to $\bar x$, we can just set $\bar x=0$ and replace 
the transition amplitude $A$ with 
$A_0=\exp(-\varepsilon W_0)/\sqrt{2\pi\varepsilon}$. 
In the remaining terms the limit has to be performed more carefully. The terms containing combinations $V_+\pm V_-$ and their derivatives 
are the simplest, and we obtain
\begin{eqnarray}
(V_++ V_-)\Big|_{\bar x\to 0}&=&2\sum_{m=0}^\infty\frac{\varepsilon^{2m}\stackrel{(2m)}{V}}{(2m)!\, 2^{2m}}\, ,\\
(V_+- V_-)\Big|_{\bar x\to 0}&=&2\sum_{m=0}^\infty\frac{\varepsilon^{2m+1}\stackrel{(2m+1)}{V}}{(2m+1)!\, 2^{2m+1}}\, ,\\
\partial (V_++V_-)\Big|_{\bar x\to 0}&=& \bar \partial (V_++V_-)\Big|_{\bar x\to 0}
= \ 2\sum_{m=0}^\infty\frac{\varepsilon^{2m}\stackrel{(2m)}{V}\hspace*{-1mm}'}{(2m)!\, 2^{2m}}\, ,
\end{eqnarray}
where the prime in the last expression denotes the derivative with respect to $x$. For the terms $\bar\partial A$ and $\partial_\tau\bar\partial A$ we have to explicitly use the full double power 
expansion for the effective potential
\begin{equation}
\label{eq:Wdouble}
W(x,\bar x;\varepsilon, \tau)=\sum_{m=0}^{\infty}\sum_{k=0}^{m}\Big[c_{m,k}(x, \tau)\,\varepsilon^{m-k}\bar x^{2k}
+ c_{m+1/2,k}(x, \tau)\,\varepsilon^{m-k}\bar x^{2k+1}\Big] \, ,
\end{equation}
perform the differentiation and take the limit afterwards. This yields the results
\begin{eqnarray}
\bar\partial A\Big|_{\bar x\to 0}&=&-\varepsilon A_0\sum_{m=0}^\infty c_{m+1/2}\,\varepsilon^{m}\, ,\\
\partial_\tau\bar\partial A\Big|_{\bar x\to 0}&=&-\varepsilon A_0\sum_{m=0}^\infty\varepsilon^m\, (\dot{c}_{m+1/2}-c_{m+1/2}\, \varepsilon\dot{W}_0)\, ,
\end{eqnarray}
where dots now represent derivatives with respect to the time argument $\tau$ of the coefficient $c_{m+1/2} \equiv c_{m+1/2, 0} $ and the effective potential $W_0$. As we see, the odd-power
 coefficients $c_{m+1/2}$ cannot be eliminated altogether, although we are considering the diagonal amplitudes, for which we have
$\bar x=0$. This is due to the derivatives with respect to 
$\bar x$. From this we can deduce 
that  we will need again two recursion relations to determine all needed coefficients, despite the fact that in the end we will use only the even-power ones. Therefore, we will use 
Eq.~(\ref{eq:dt}) to derive the second recursive relation for the coefficients. In order to do so, we still need to calculate the term $\partial\bar\partial A$ in the considered limit $\bar x\to 0$:
\begin{equation}
\partial\bar\partial A\Big|_{\bar x\to 0}=A_0\sum_{m=0}^\infty\varepsilon^m\, (c_{m+1/2}\, \varepsilon^2 W'_0 -c'_{m+1/2}\, \varepsilon) \, .
\end{equation}
Inserting all calculated $\bar x\to 0$ terms into equations (\ref{eq:dt}) and (\ref{eq:A0comb}), as well as using the expansion (\ref{eq:W0double}), we finally obtain the following coupled system of 
recursive relations for the coefficients $c_m$ and $c_{m+1/2}$:
\begin{eqnarray}
&&\hspace*{-5mm}(2m+1)\, c'_m=\Pi(0, m)\, \frac{\stackrel{(m)}{V}\hspace*{-1mm}'}{m!\, 2^m}+ \frac{1}{4}\,c_{m-1}'''+\frac{1}{2}\dot{c}_{m-1/2}-2\sum_l\frac{\stackrel{(2l)}{V}}{(2l)!\, 2^{2l}}\, 
c'_{m-2l-1}\nonumber\\
&&\hspace*{10mm}+ \sum_l\frac{\stackrel{(2l+1)}{V}}{(2l+1)!\, 2^{2l+1}}\, c_{m-2l-3/2}
+2\sum_l c'_l\, c_{m-l-1}+ 2\sum_l l\, c_l\, c'_{m-l-1}\nonumber\\
&&\hspace*{10mm}-\frac{3}{4}\sum_l c'_l\, c''_{m-l-2}-\frac{1}{2}\sum_l c_{l+1/2}\, \dot{c}_{m-l-2}
+\frac{1}{4}\sum_{l, r} c'_l\, c'_r\, c'_{m-l-r-3}\, ,
\label{eq:cmrec}\\
\label{eq:cm12rec}
&&\hspace*{-5mm}\frac{1}{2}\, c'_{m+1/2}=-2\, \Pi(0, m)\, \frac{\stackrel{(m+1)}{V}}{(m+1)!\, 2^{m+1}}+\dot{c}_m+\frac{1}{2}\sum_l c_{l+1/2}\, c'_{m-l-1}\, .
\end{eqnarray}

The above recursive relations are solved in a similar way as before. In order to obtain the level $p$ diagonal effective action $W_0$, we need to take into account the terms in the expansion with 
$m=0,1,\ldots ,p-1$. The recursions for $c'_m$ and $c'_{m+1/2}$ are easily solved starting from $m=0$ up to a desired level $p-1$. Although we get in this way
only their first derivatives with respect to $x$, the 
coefficients themselves 
can be calculated by direct symbolic integration, and all solutions can be expressed in a closed form. The explicit calculation of the coefficients to high orders yields the same results 
we have obtained in the previous section. To order $p=1$, we only have the trivial equation
\begin{equation}
c'_0=V'\, ,
\end{equation}
that gives the well-known boundary condition $c_0=V$. To order $p=2$ we have
\begin{eqnarray}
&&\hspace*{-10mm}
c'_{1/2}=-2\dot{V}+2\dot{c}_0=0\, ,\\
&&\hspace*{-9mm}
3c'_1=\frac{1}{4}c'''_0+\frac{1}{2}\dot{c}_{1/2}-2Vc'_0+2c'_0\, c_0=\frac{1}{4}c'''_0\, ,
\end{eqnarray}
which yields $c_{1/2}=0$ and $c_1=c''_0/12=V''/12$. To order $p=3$ we first calculate the odd-power coefficient,
\begin{equation}
c'_{3/2}=2\dot{c}_1+c_{1/2}\, c'_0=\frac{1}{6}\dot{c}''_0\, ,
\end{equation}
which leads to the result $c_{3/2}=\dot{c}'_0/6=\dot{V}'/6$.  The even-power coefficient $c_2$ is obtained from
\begin{eqnarray}
5c'_2&=&\frac{1}{8}\ddot{V}'+\frac{1}{4}c'''_1+\frac{1}{2}\dot{c}_{3/2}-2Vc'_1+\frac{1}{2}\dot{V}c_{1/2}+4c'_0\, c_1
+2c'_1\, c_0-\frac{1}{2}c_{1/2}\, \dot{c}_0-\frac{3}{4}c'_0\, c''_0\nonumber\\
&=&\left(\frac{5\ddot{V}}{24} + \frac{V^{(4)}}{48}-\frac{5V'^2}{24}\right)'\, ,
\end{eqnarray}
which finally yields
\begin{equation}
c_2=\frac{\ddot{V}}{24} + \frac{V^{(4)}}{240}-\frac{V'^2}{24}\, .
\end{equation}
These results coincide with the results obtained in our earlier paper \cite{fcpitdp1} from the general recursion relations. We similarly proceed to calculate higher-order coefficients. This is easily automated in symbolic calculus software packages like Mathematica \cite{mathematica}, and we have implemented the derived recursions as a part of our SPEEDUP \cite{speedup} code.

The main advantage of this approach is that we have been able to derive recursion relations involving only the lower-level coefficients $c_m=c_{m, 0}$ and $c_{m+1/2}=c_{m+1/2, 0}$, thus not requiring the 
calculation of all even- and odd-power coefficients, which are needed for the general case. At the end of this section, we note that for time-independent potential $V$ the recursive relation 
(\ref{eq:cmrec}) reduces to the previously known result \cite{balazpre}, while the second recursion (\ref{eq:cm12rec}) yields the expected result $c_{m+1/2}=0$. Note that our 
recursive approach thus allows to calculate higher orders of the seminal Wigner expansion \cite{wigner}.

\begin{figure}[!t]
\centering
\includegraphics[width=8.1cm]{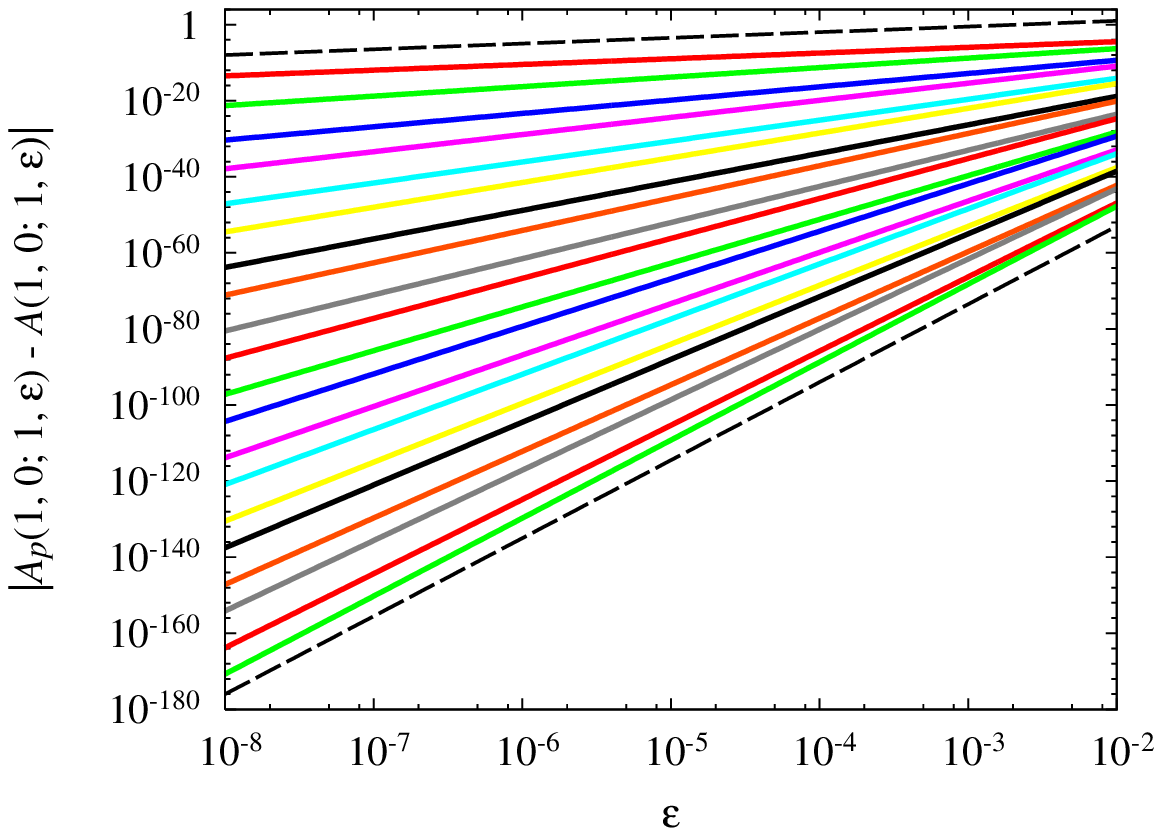}
\includegraphics[width=8.1cm]{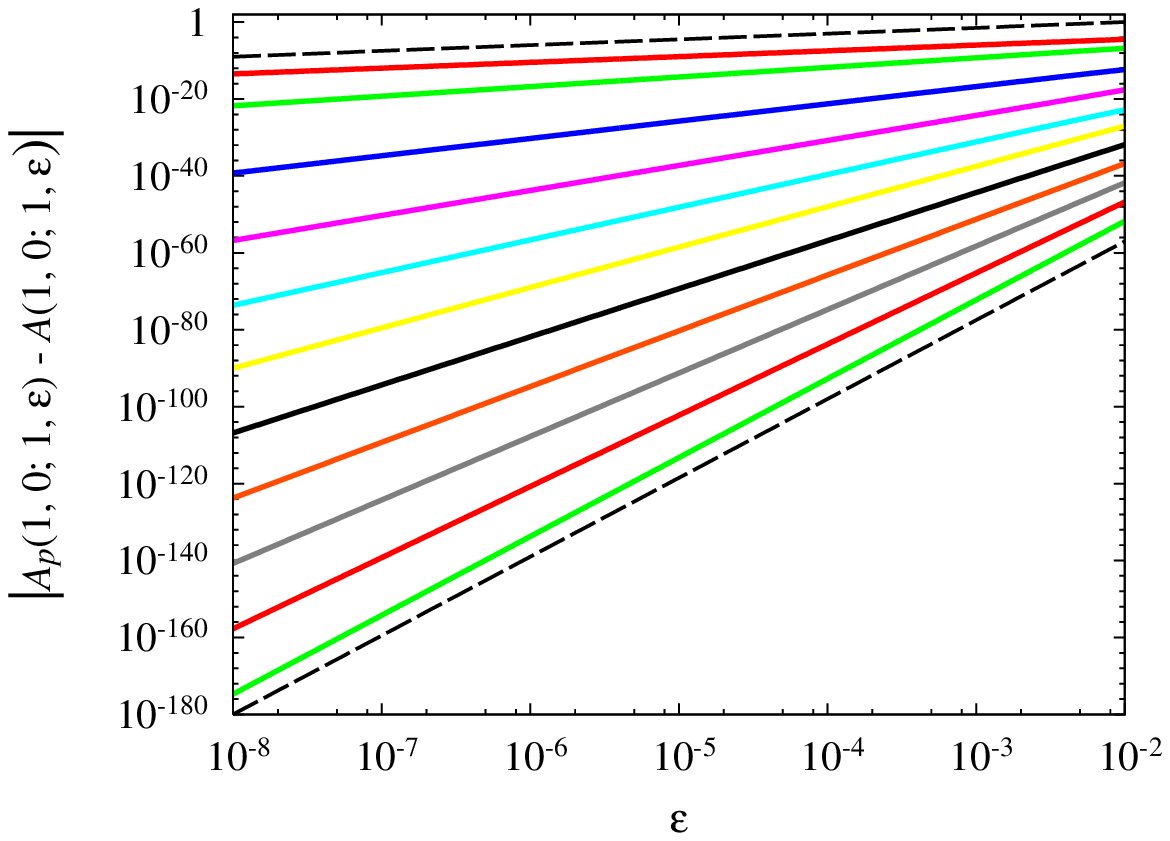}
\caption{Deviations of diagonal transition amplitudes $|A_p(1, 0; 1, \varepsilon)-A(1, 0; 1, \varepsilon)|$ as a function of propagation time $\varepsilon$ for: (left) 
time-dependent harmonic oscillator (\ref{eq:hogrosche}),  calculated analytically for   $p=1, 2, 3, 4, \ldots, 20$ from top to bottom; (right) forced harmonic oscillator (\ref{eq:fho}) with $\omega=1$ 
and $\Omega=2$, calculated analytically for  $p=1, 2, 4, 6, 8, 10, 12, 14, 16, 18, 20$ from top to bottom. 
The dashed lines on both graphs are proportional to $\varepsilon^{1.5}$ and $\varepsilon^{20.5}$, 
and  demonstrate the perfect scaling of the 
corresponding level $p=1$ and $p=20$ results.}
\label{fig:diag}
\end{figure}

Fig.~\ref{fig:diag} illustrates practical advantages of using velocity-independent effective actions for the numerical calculation of diagonal transition amplitudes. The plot on the left gives the 
deviations of diagonal amplitudes calculated with different levels $p$ of the effective potential $W_0$ for the Grosche-rescaled \cite{groschepla} harmonic oscillator,
\begin{equation}
\label{eq:hogrosche}
V_{\rm G, HO}(x, t)=\frac{\omega^2 x^2}{2 (1+t^2)^2}\, .
\end{equation}
 while the plot on the right gives the analogous results for the forced harmonic oscillator
\begin{equation}
\label{eq:fho}
V_{\rm FHO}(x, t)=\frac{1}{2}\omega^2x^2-x\sin\Omega t\, ,
\end{equation}
where $\Omega$ denotes the frequency of the external driving field. Both models are exactly solvable, and the obtained $\varepsilon$-scaling to exceedingly high orders $p$ demonstrates the correctness of the analytically derived results.

\section{Real-time formalism}
\label{sec:rt}
The presented approach has so far been developed within the imaginary-time framework, which is useful in many practical applications. However, in order to study 
the more relevant real-time 
dynamics of quantum systems, we have to switch back to the real-time formalism. One possibility would be to make all calculations in imaginary time, and then to try to perform an 
inverse Wick rotation, which might be difficult due to the oscillatory nature of the integrand in calculating the real-time propagator. Another, much more straight-forward possibility is to derive a 
new set of recursive relations within the real-time formalism. In this section we will briefly outline such a procedure.

Reverting the imaginary-time formalism into a real-time one is achieved by replacing the variable $t$ representing the imaginary time with $i t_\mathrm{R}$ in all 
expressions, where now $t_\mathrm{R}$ represents the real time. This includes also the replacement of the time-interval $\varepsilon$ with $i\varepsilon_\mathrm{R}$, and the 
time-midpoint $\tau$ with its real-time counterpart $i\tau_\mathrm{R}$. Instead of (\ref{eq:Aexacta}), the short-time transition amplitude is now expressed as
\begin{equation}
\label{eq:Aexactart}
A(\mathbf a, t_a; \mathbf b,t_b)=\frac{1}{(2\pi i \varepsilon_\mathrm{R})^{Pd/2}}\, e^{i S^*(\mathbf x, \bar{\mathbf x}; \varepsilon_\mathrm{R}, \tau_\mathrm{R})}\, ,
\end{equation}
and the real-time version of the ideal effective action is defined as
\begin{equation}
\label{eq:Aexactbrt}
S^*(\mathbf x, \bar{\mathbf x}; \varepsilon_\mathrm{R}, \tau_\mathrm{R})
= \frac{2}{\varepsilon_\mathrm{R}}\bar{\mathbf x}^2-\varepsilon_\mathrm{R} W(\mathbf x, \bar{\mathbf x}; \varepsilon_\mathrm{R}, \tau_\mathrm{R})\, ,
\end{equation}
which represents the counterpart of Eq.~(\ref{eq:Aexactb}). Following the same procedure outlined in Sec.~\ref{sec:eff}, we arrive at the real-time counterpart of Eq.~(\ref{eq:Weq}) for the effective potential,
\begin{equation}
W+\bar{\mathbf x}\cdot\bar{\boldsymbol{\partial}}\,W+\varepsilon\partial_{\varepsilon} W
-\frac{i}{8}\,\varepsilon\partial^2 W-\frac{i}{8}\,\varepsilon\bar\partial^2 W
-\frac{1}{8}\,\varepsilon^2(\boldsymbol{\partial} W)^2
-\frac{1}{8}\,\varepsilon^2(\bar{\boldsymbol{\partial}} W)^2=\frac{1}{2}\, (V_++V_-)\, ,
\label{eq:Weqrt}
\end{equation}
where the subscript R is dropped for simplicity. Further derivation of real-time recursion relations is a straight-forward task. For brevity, we will not give the explicit 
form of the recursion relations, but their Mathematica implementation is available from the SPEEDUP code web page \cite{speedup}.

We illustrate the applicability of this formalism for studying the real-time dynamics 
within the space-discretized approach \cite{diag1, diag2}. If we discretize the continuous 
space and replace it with a grid defined by a discretization step $\Delta$, all quantities are only defined on a discrete set of coordinates $\mathbf{q}_\mathbf{n}=\mathbf{n}\Delta$, where 
$\mathbf{n}\in\mathbb{Z}^{Pd}$ is a vector of $Pd$ integer numbers. Matrix elements of the evolution operator,
\begin{equation}
U_{\mathbf{nm}}(t_a \rightarrow t_b)=\langle \mathbf{q_m} | \hat U(t_a\rightarrow t_b) | \mathbf{q_n}\rangle\, ,
\end{equation}
represent real-time transition amplitudes $A_\mathbf{nm}(t_a, t_b) = A(\mathbf{q_n}, t_a; \mathbf{q_m}, t_b)$, which can be calculated using the real-time effective action approach. If the initial 
state of the system $|\psi, t_a\rangle$ is represented by a vector $\psi(t_a)$ whose elements are $\psi_\mathbf{n}(t_a)=\psi(\mathbf{q_n}, t_a) = \langle \mathbf{q_n} | \psi, t_a\rangle$, its dynamics 
can be calculated by a simple matrix multiplication $\psi(t_b)=U(t_a \rightarrow t_b) \cdot \psi(t_a)$, i.e.
\begin{equation}
\psi_\mathbf{n}(t_b)=\sum_\mathbf{m} U_\mathbf{nm}(t_a \rightarrow t_b)\, \psi_\mathbf{m}(t_a)\, .
 \end{equation}
Therefore, since the matrix elements of the evolution operator can be accurately calculated using the effective action approach, we are able to study real-time dynamics of 
the system starting from any desired initial state.

Note that, although we rely on the short-time expansion of transition amplitudes, we are not limited to study only a short-time evolution, since the above matrix multiplication 
can be repeatedly performed. For any given propagation time $T$, we can divide the evolution to $N$ sub-intervals of length $\varepsilon=T/N$, which now correspond to short-time 
evolution matrix elements $U_\mathbf{nm}(\varepsilon)$. In addition to this, using the higher-order effective actions makes it possible to perform
high-accuracy calculations of $U_\mathbf{nm}$, and, correspondingly, to eliminate the associated numerical errors for all practical purposes.

\begin{figure}[!t]
\centering
\includegraphics[width=8.1cm]{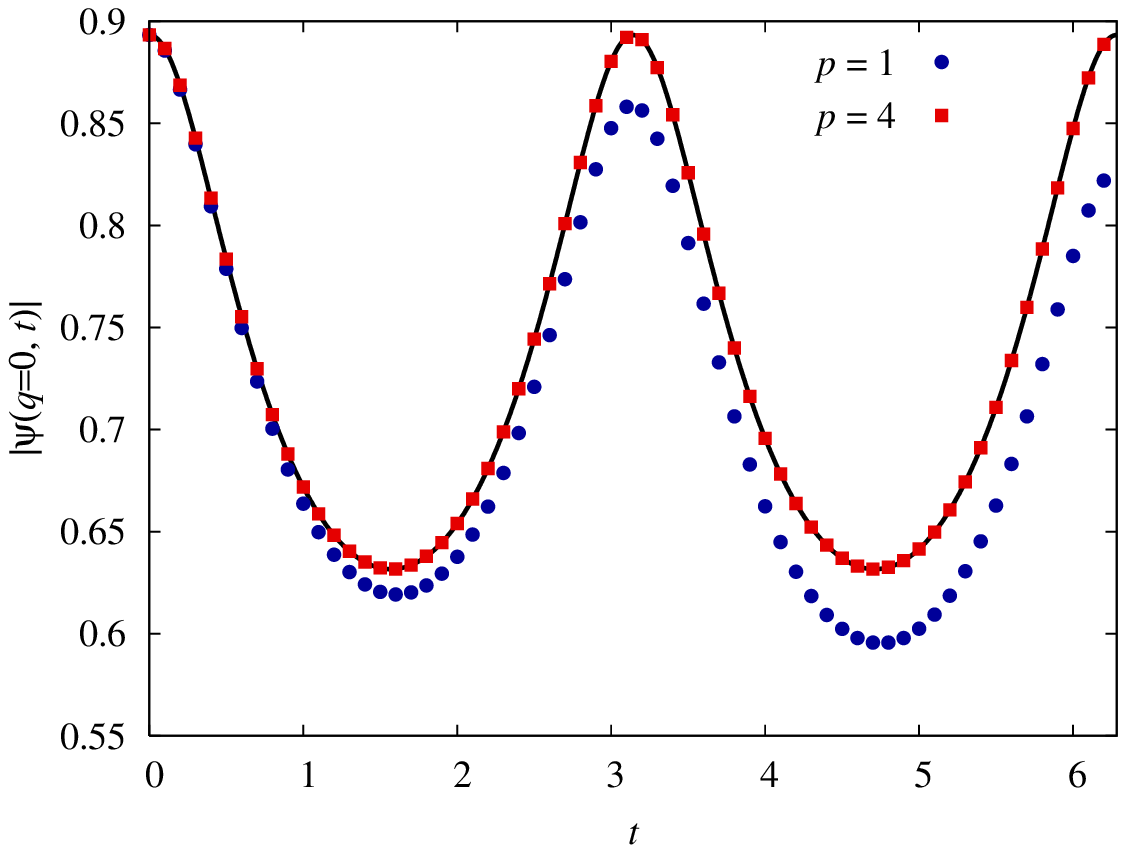}
\includegraphics[width=8.1cm]{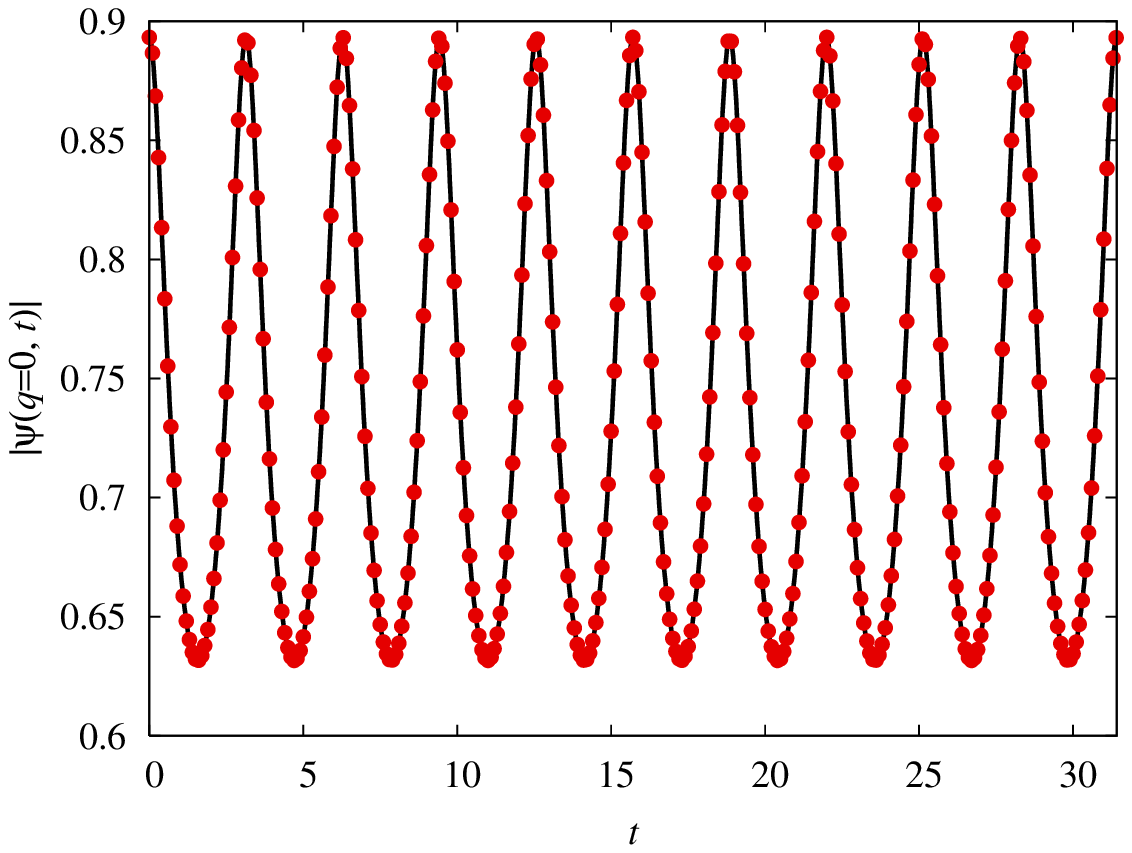}
\caption{Time evolution of the one-dimensional harmonic oscillator $V(q)=\frac{1}{2}\,\omega^2 q^2$ calculated using the space-discretized method 
\cite{diag1, diag2} and the effective action approach with: (left) $p=1$, $4$ and (right) $p=20$. Both graphs display the time dependence $|\psi(q=0, t)|$ of the absolute value of 
the wave function at $q=0$, and solid line represents the exact solution. The harmonic frequency was $\omega=1$, the time-interval for propagation was
$\varepsilon=0.1$, and the initial state was set to the ground state of the harmonic oscillator with $\omega=2$.}
\label{fig:rt1}
\end{figure}

\begin{figure}[!t]
\centering
\includegraphics[width=12cm]{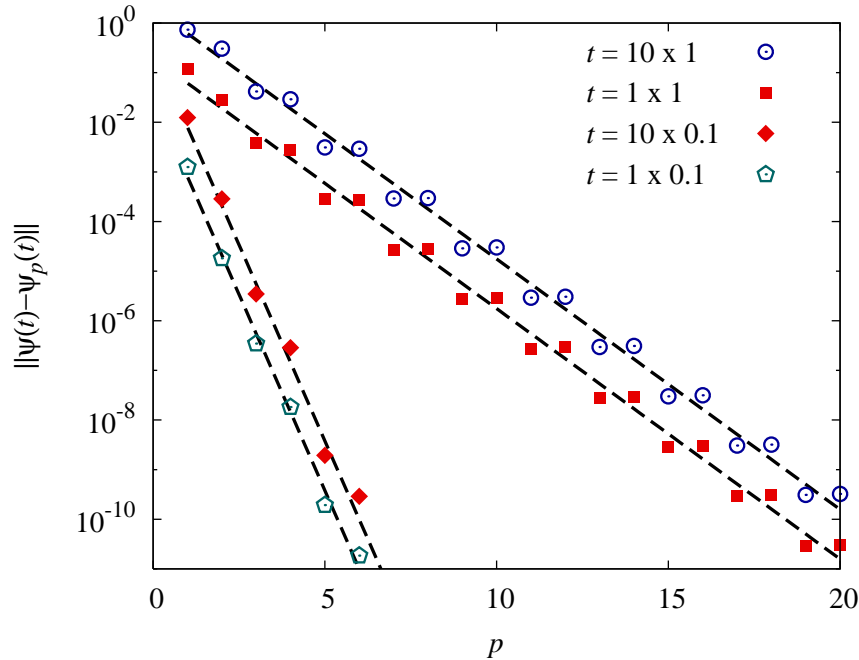}
\caption{Integral measure (\ref{eq:im}) for numerically calculated time evolution of the harmonic oscillator as a function of the effective action level $p$ for different values of the propagation time $t$. The parameters are the same as in Fig.~\ref{fig:rt1}.}
\label{fig:rt2}
\end{figure}

To demonstrate this,  we show in Fig.~\ref{fig:rt1} the
time evolution of a harmonic oscillator $V(q)=\frac{1}{2}\,\omega^2 q^2$ calculated using the described method with effective action levels $p=1$, $4$ and $p=20$. As we can see, using the propagation interval $\varepsilon=0.1$, the naive $p=1$ action can be used only for short propagations times, while we are able to reproduce very accurately the long-time evolution of the system with higher $p$ levels. In order to further quantitatively assess numerical errors of the obtained results, we use the following integral measure,
\begin{equation}
||\psi(t)-\psi_p(t)||=\left(\int_{-\infty}^\infty |\psi(q, t)-\psi_p(q, t)|^2\, dq\right)^{1/2}\, ,
\label{eq:im}
\end{equation}
where $\psi(q, t)$ represents the exact time evolution of the wave function and $\psi_p(q, t)$ is the approximate time evolution calculated using level $p$ effective action. 
The semi-log plot in Fig.~\ref{fig:rt2} gives the $p$-dependence of the above-defined integral measure and demonstrates that it obeys the expected power law, i.e.
$\varepsilon^{p+1/2}$ in this case, leading to a much smaller error when the propagation interval is reduced from $\varepsilon=1$ to $\varepsilon=0.1$. This graph also shows 
that errors due to the repeated matrix multiplication accumulate linearly with the number of time steps.

The study of errors presented in Fig.~\ref{fig:rt2} is very instrumental in optimizing numerical parameters in practical applications. If we compare errors, which 
correspond to the same total evolution time $t=1$ and are calculated for a 
propagation in one time-step $\varepsilon=t$ and in $N=10$ steps $\varepsilon=t/N$, we can see that decreasing 
$\varepsilon$ substantially reduces errors. This is easily understood, since errors are proportional to $\varepsilon^{p+1-Pd/2}$ and, therefore, introducing $N$ time steps 
is expected to reduce errors by a factor of $N^{p+1-Pd/2}$. However, the fact that the 
matrix multiplication will have to be repeated $N$ times introduces an additional factor of $N$, 
thus leading to the total reduction factor of $N^{p-Pd/2}$. 

\begin{figure}[!t]
\centering
\includegraphics[width=12cm]{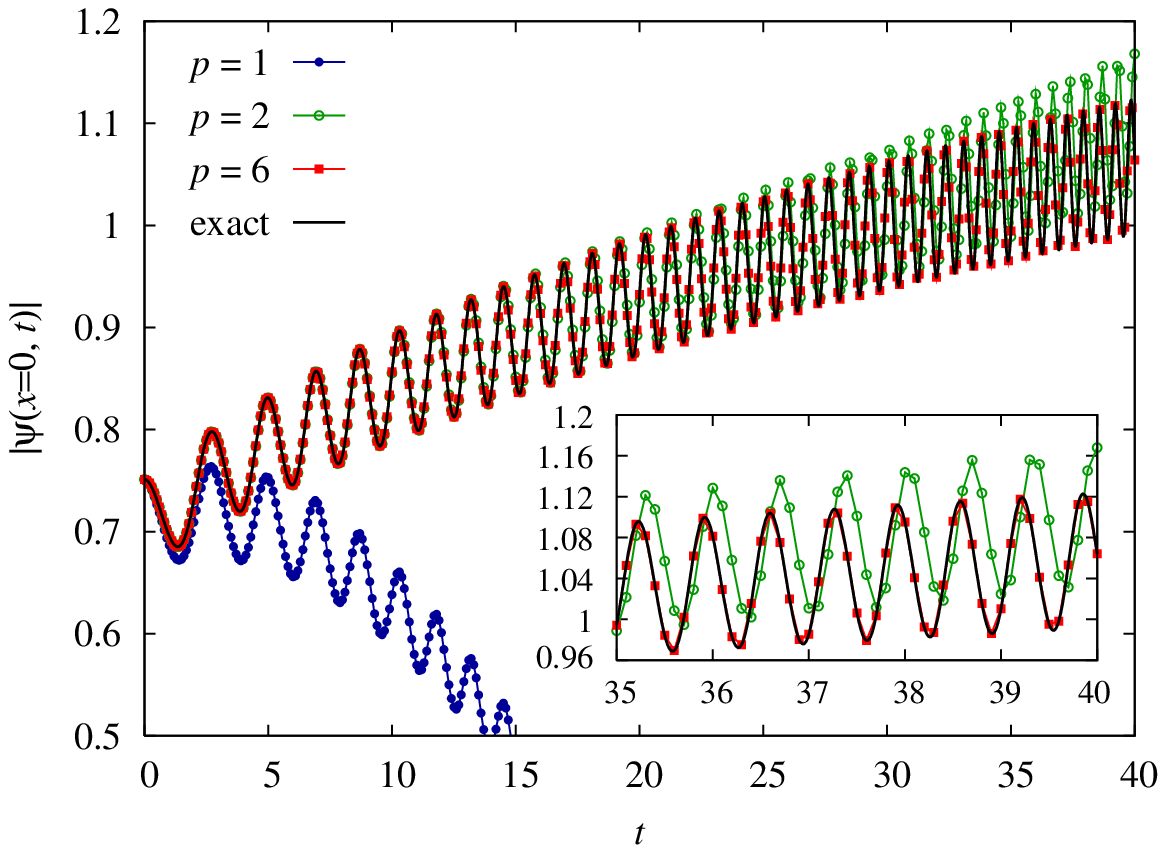}
\caption{Time evolution of the time-dependent harmonic oscillator (\ref{eq:tdho}) calculated using the space-discretized method \cite{diag1, diag2} and the effective action approach with $p=1$, $p=2$, and $p=6$. The graph displays the time dependence $|\psi(q=0, t)|$ of the absolute value of the wave function at $q=0$. The time-dependent harmonic frequency is given by $\omega(t)=1+\frac{1}{10}\, t$, time-interval for propagation was $\varepsilon=0.1$, and the initial state was set according to Eq.~(\ref{eq:tdhopsi0}).}
\label{fig:rt3}
\end{figure}

As a final example, we calculate the time evolution of the time-dependent harmonic oscillator with the potential
\begin{equation}
\label{eq:tdho}
V(q, t)=\frac{1}{2}\, \omega^2(t) q^2\,
\end{equation}
with the frequency $\omega(t)=1+\frac{1}{10}\, t$ for $p=1$ and $p=6$. Fig.~\ref{fig:rt3} displays the time evolution of the absolute value of the wave function at $q=0$ with the propagation interval $\varepsilon=0.1$, and the initial state set to
\begin{equation}
\label{eq:tdhopsi0}
\psi(q, t=0)=\frac{1}{\pi^{1/4}}\, e^{-\frac{1}{2}q^2+\frac{1}{2}iq}\, .
\end{equation}
As expected, the naive $p=1$ effective action can only be used for very short propagation times, while $p=2$ action gives accurate results for longer propagation times $T\leq 15$. A moderate level $p=6$ effective action represents an even further substantial improvement and can be used to accurately study much longer propagation times, as can be seen from the inset in Fig.~\ref{fig:rt3}.

We emphasize that the presented approach might be especially of interest for Path Integral Monte Carlo (PIMC) calculations of the dynamics of quantum systems, in conjunction e.g. with the multilevel blocking method \cite{mak}. Dynamical PIMC calculations are quite difficult due to the dynamical sign problem, and the availability of highly accurate propagators allows use of relatively small number of Trotter slices, thus substantially improving the numerical convergence.

\section{Conclusions}
\label{sec:conclusions}
In this paper we have presented a significant extension of the approach introduced in the preceding paper \cite{fcpitdp1}, which has established a recursive procedure for calculating the short-time transition amplitudes for one-dimensional quantum systems in time-dependent potentials. This approach is generalized here to non-relativistic many-body quantum systems with many degrees of freedom. In parallel to the approach for time-independent potentials \cite{balazpre}, we have introduced an ideal effective potential for time-dependent systems, derived the appropriate equation using the forward and backward Schr\" odinger equation for the transition amplitude, and set up an efficient system of recursive relations, which can be analytically solved to high orders in the short propagation time. Furthermore, we have implemented a symbolic calculation scheme for higher-order effective actions in the SPEEDUP code \cite{speedup}. The analytically derived results are verified by studying several models and a list of possible applications of the presented method to relevant many-body quantum systems has been briefly outlined.

In addition to this, we have also studied velocity-independent part of the effective action, which is relevant for calculating the diagonal amplitudes and partition functions. We have obtained a new, simpler set of recursion relations, which determine the diagonal effective action, and have numerically verified that it yields the correct systematic increase in the convergence of diagonal amplitudes for several models.

Finally, we have also looked at how the developed formalism can be transformed from its original, imaginary-time setup to the real-time one. We have derived the real-time counterparts of equations for the effective potential and applied the higher-order real-time effective actions to a numerical study of the time evolution of several models using the space-discretized approach \cite{diag1, diag2}. We have demonstrated that the presented approach can be successfully used both in the real-time and in the imaginary-time formalism, and that in both cases we obtain analytically derived improved convergence of numerically calculated transition amplitudes and other quantities. We point out that the presented approach can contribute to improving Path Integral Monte Carlo calculations of the dynamics of quantum systems in conjunction e.g. with the multilevel blocking method, since higher-order propagators enable use of fewer number of Trotter slices.

\section*{Acknowledgements}
We thank Hagen Kleinert for several useful suggestions.
This work was supported in part by the Ministry of Science and Technological Development of the Republic of Serbia, under project No. ON171017 and bilateral projects PI-BEC and NAD-BEC, funded jointly with the German Academic Exchange Service (DAAD). Numerical simulations were run on the AEGIS e-Infrastructure, supported in part by FP7 project EGI-InSPIRE, HP-SEE and PRACE-1IP.

\end{document}